\documentclass[twocolumn]{aastex62}

\newcommand{\HII}{H\tiny{ }\footnotesize{II}\normalsize{ }}
\newcommand{\NII}{[N\tiny{ }\footnotesize{II}\normalsize{] }}
\newcommand{\SII}{[S\tiny{ }\footnotesize{II}\normalsize{] }}
\newcommand{\NIIs}{[N\tiny{ }\footnotesize{II}\normalsize{]}}
\newcommand{\SIIs}{[S\tiny{ }\footnotesize{II}\normalsize{]}}
\newcommand{\OIII}{[O\tiny{ }\footnotesize{III}\normalsize{] }}
\newcommand{\OIIIs}{[O\tiny{ }\footnotesize{III}\normalsize{]}}

\newcommand{\Ha}{H$\alpha$ }
\newcommand{\Has}{H$\alpha$}

\defcitealias{Dopita16}{D16}

\graphicspath{{./}{figures/}}

\submitjournal{ApJ}

\shorttitle{Extending the MZR to low mass at high-$z$}
\shortauthors{Cameron et al.}

\begin{document}


\title{Prospects for extending the Mass-Metallicity Relation to low mass at high redshift: a case study at $z\sim1$}

\correspondingauthor{Alex J. Cameron}
\email{alexc@student.unimelb.edu.au}

\author[0000-0002-0450-7306]{Alex J. Cameron}
\affil{School of Physics, The University of Melbourne, Parkville, VIC 3010, Australia}
\affil{ARC Centre of Excellence for All Sky Astrophysics in 3 Dimensions (ASTRO 3D), Australia}

\author{Tucker Jones}
\affiliation{Department of Physics, University of California, Davis, 1 Shields Avenue, Davis, CA 95616, USA}

\author{Tiantian Yuan}
\affiliation{Centre for Astrophysics and Supercomputing, Swinburne University of Technology, Hawthorn, Victoria 3122, Australia}
\affil{ARC Centre of Excellence for All Sky Astrophysics in 3 Dimensions (ASTRO 3D), Australia}

\author{Michele Trenti}
\affil{School of Physics, The University of Melbourne, Parkville, VIC 3010, Australia}
\affil{ARC Centre of Excellence for All Sky Astrophysics in 3 Dimensions (ASTRO 3D), Australia}

\author{Stephanie Bernard}
\affil{School of Physics, The University of Melbourne, Parkville, VIC 3010, Australia}

\author{Alaina Henry}
\affil{Space Telescope Science Institute, 3700 San Martin Drive
Baltimore, MD 21218, USA}

\author{Austin Hoag}
\affil{Department of Physics and Astronomy, University of California, Los Angeles, 430 Portola Plaza, Los Angeles, CA 90095, USA}

\author{Benedetta Vulcani}
\affil{INAF--Osservatorio astronomico di Padova, Vicolo Osservatorio 5, I-35122 Padova, Italy}



\begin{abstract}

We report $J$-band MOSFIRE spectroscopy of a low-mass ($\text{log}(M_*/M_\odot)=8.62^{+0.10}_{-0.06}$) star-forming galaxy at $z=0.997$ showing the detection of \NII and \SII alongside a strong \Ha line. We derive a gas-phase metallicity of $\text{log}(\text{O}/\text{H})=7.99^{+0.13}_{-0.23}$, placing this object in a region of $M_* - Z$ space that is sparsely populated at this redshift. Furthermore, many existing metallicity measurements in this $M_* - z$ regime are derived from only \NIIs/\Ha (N2), a diagnostic widely used in high-redshift metallicity studies despite the known strong degeneracy with the ionization parameter and resulting large systematic uncertainty. We demonstrate that even in a regime where \NII and \SII are at the detection limit and the measurement uncertainty associated with the \NIIs/\SII ratio is high ($S/N\approx3$), the more sophisticated Dopita et al. diagnostic provides an improved constraint compared to N2 by reducing the systematic uncertainty due to the ionization parameter.
This approach does not, however, dispel uncertainty associated with stochastic or systematic variations in the nitrogen-to-oxygen abundance ratio. While this approach improves upon N2, future progress in extending metallicity studies into this low-mass regime will require larger samples to allow for stochastic variations, as well as careful consideration of the global trends among dwarf galaxies in all physical parameters, not just metallicity.

\end{abstract}



\section{Introduction} \label{sec:intro}

Measurements of the gas-phase oxygen abundance (metallicity hereafter) in galaxies provide powerful insights into the galaxy-scale star-formation and gas-flow processes that have shaped the development of the galaxy population across cosmic time.
Despite extensive studies into the tight correlation between metallicity and stellar mass in galaxies, so-called Mass-Metallicity Relation (MZR), much debate still exists as to its origin \citep{Lequeux79, Skillman89, Tremonti04, Berg12, Andrews13, Yabe14, Sanders15, Maiolino19}.
The MZR is observed to have evolved with redshift, with lower average metallicities observed at earlier cosmic times for a given stellar mass \citep{Savaglio05, Erb06, Maiolino08}. However, constraints on the evolution of shape and scatter of the MZR are less clear, due in particular to the difficulties associated with making individual measurements of low-mass galaxies at high redshift.
While average metallicity evolution is an important input into galaxy evolution models, extending constraints on chemical evolution to lower mass objects promises key insights into the evolution of the galaxy population.

Existing studies suggest that the shape of the MZR is not constant across cosmic time \citep{Zahid13, Zahid14b}.
These studies support a downsizing scenario in which low-mass galaxies enrich onto the local MZR at later times.
However, despite predictions from theory suggesting that measuring the shape and scatter of the MZR below $\text{log}(M_*/M_\odot)<9.0$ provide the best prospects for disentangling the driving forces behind this evolution \citep[e.g.][]{Dave12}, high-redshift studies rarely extend into this mass regime.
Thus, the MZR is poorly constrained for low-mass galaxies at high-redshift ($z\gtrsim1$) as these observations are difficult to carry out, leading to small sample sizes of often loosely constrained measurements \citep{Zahid11, Wuyts12, Wuyts14, Wuyts16, Belli13, Henry13a, Henry13b, Yuan13, Amorin14, Maseda14}.
Larger samples of low-mass measurements are required to disentangle the impact of different processes on the evolution of galaxies across cosmic time.

An additional challenge in studies of the MZR below $\text{log}(M_*/M_\odot)<9.0$ lies with uncertainties in how the metallicities are derived.
Diagnostics based on electron temperature ($T_e$) are widely considered the most reliable measures of metallicity \citep[e.g., review by][]{Maiolino19}.
However the range of application of this ``direct'' method is limited to objects in which the weak \OIII $\lambda4363$ emission line can be observed, making it unfeasible beyond moderate redshifts \citep{Jones15, Ly16a, Ly16b, Calabro17}.
Motivated by these difficulties, numerous diagnostics have been developed based on ratios of the most easily detected strong rest-frame optical emission lines, calibrated from stellar population synthesis and photoionization models \citep[e.g.][]{KewleyDopita02} or $T_e$ measurements taken in either the local universe \citep[e.g.][]{PettiniPagel04} or at moderate redshifts \citep{Jones15}.
Alternatively, a number of generalized approaches exist that compare a range of strong-line fluxes to photoionization models to simultaneously fit for metallicity alongside other key physical parameters \citep[e.g.][]{PerezMontero14, Blanc15, ValeAsari16}.
As metallicity studies are extended to higher redshifts and lower masses, these strong-line methods become an essential tool in understanding the galaxy population.

Strong-line diagnostics greatly extend the range of stellar masses and redshifts over which metallicities can be derived, however questions remain about their reliability \citep{KewleyEllison08, Steidel14}.
In particular, when the number of observed emission lines is small, strong-line diagnostics often fail to disentangle the degeneracy between metallicity and the effects of other physical parameters such as ionization parameter, electron density, hardness of the ionizing sources and relative abundance ratios \citep[e.g.][]{MoralesLuis14,Maiolino19}.
These issues are omnipresent in studies targeting low-mass galaxies at high-redshift where observational challenges frequently limit the range of available emission lines, meaning metallicities are often derived simply from \NIIs/\Ha (N2; \citealt{PettiniPagel04}).

The N2 ratio has its advantages in that the two lines are close in wavelength, such that they can be obtained in a single exposure and the ratio is independent of reddening. Thus it has been useful in expanding measurements of the metallicity to faint galaxies \citep[e.g.][]{Erb06, Yabe14}.
However, metallicities derived from this line ratio contain large systematic uncertainties due to strong degeneracy with ionization parameter.
Additionally this ratio is primarily sensitive to nitrogen abundance, whereas oxygen abundance is derived with some assumed N/O ratio, introducing further uncertainty \citep[e.g.][]{PerezMontero09, PerezMontero14}.

In general, measuring larger suites of emission lines will likely be critical to provide robust metallicity measurements, thereby improving constraints on the chemical evolution of galaxies.
In line with this, \citet{Dopita16} (\citetalias{Dopita16} hereafter) have proposed that the set of H$\alpha$, \NII $\lambda$6584 and \SII $\lambda\lambda$6717, 6731 rest-frame optical lines will prove convenient in this pursuit.
The relatively narrow wavelength range covered by H$\alpha$, \NII and \SII lends itself kindly to high-redshift studies as derived line ratios are almost independent of reddening and can typically be observed in one spectroscopic exposure.
Additionally, provided the \NIIs/\SII line ratio can be adequately constrained, systematic variation in the derived metallicity caused by degeneracy with ionization parameter and interstellar medium (ISM) pressure is reduced, significantly improving uncertainty as compared to methods utilizing N2 in isolation.

In this contribution, we report $J$-band spectroscopy taken with \emph{Keck}/MOSFIRE of a low-mass ($\text{log}(M_*/M_\odot)\sim 8.6$) star-forming galaxy at $z\sim1$ covering the \NII and \SII emission lines. With a moderate integration time, we achieve an improved constraint on metallicity using the \citetalias{Dopita16} diagnostic compared to a diagnostic based on N2 alone (once systematic uncertainty is considered). In addition, the combination of ground and space-based spectroscopy covers a suite of line flux measurements that is unique for a $z\sim1$ dwarf galaxy, allowing us to test consistency of a handful of metallicity diagnostics. Based on this finding, we suggest that targeted surveys utilizing existing cutting-edge instruments could leverage this diagnostic to place powerful constraints on the processes that govern the evolution of galaxies across cosmic time.

The paper is structured as follows. In Section \ref{sec:data} we provide details on the collection of the near-infrared spectral data. Section \ref{sec:analysis}  describes our analysis of the data.  Section \ref{sec:discussion} presents a brief discussion of the results before we sum up in section \ref{sec:conclusion}. Throughout this letter we adopt the \citet{Planck16} cosmology: $\Omega_\Lambda = 0.692$, $\Omega_M = 0.308$, $\sigma_8 = 0.815$, and $H_0 = 67.8$ km s$^{-1}$ Mpc$^{-1}$. All magnitudes are quoted in the $AB$ magnitude system \citep{Oke&Gunn83}. Unless otherwise stated, \NII and \SII refer to \NII $\lambda$6584 and \SII $\lambda\lambda$6717, 6731 respectively.

\section{Data} \label{sec:data}

MACS0744\_667.0 is a star-forming galaxy at $z=0.997$ with $m_{\text{AB}} = 23.35\pm0.02$ in $J$-band (\emph{HST}/WFC3 F125W) and $R_{\text{eff}}=2.49$ kpc, magnified 1.4$\times$ by cluster MACS0744 (lens modeling from \citealt{Hoag19}). The source was selected from the HST Grism Lens-Amplified Survey from Space \citep[GLASS;][]{TreuGLASS} as an intermediate-redshift target for spectroscopic follow-up with MOSFIRE at Keck (program \#Z045M, PI Trenti). Observations were carried out on March 20th, 2016 under good seeing conditions ($\sim$ 0\farcs4-0\farcs7 in J band), low atmospheric attenuation ($\Delta m < 0.1)$, and minimal airmass (1.05-1.15) for a total of 8457s, divided into individual exposures of 120s each. An ABBA dither pattern with 3\arcsec~nodding along the slit was employed and observations started at 19:35HST after acquisition of a standard star during twilight. The MOSFIRE mask included two stars ($m_J=16.1$ and $m_J=16.3$) inside the GLASS HST/WFC3 field of view, that were used to verify source alignment.

The MOSFIRE data were reduced using the publicly available data reduction pipeline (DRP\footnote{\url{https://www2.keck.hawaii.edu/inst/mosfire/drp.html}}). The DRP performs wavelength calibration, rectification, background subtraction and skyline subtraction for each 2D slit in the multi-object slit-mask. The resulting outputs of the DRP are individual 2D signal and noise spectra in electrons per second for each slit on the mask.


\begin{figure*}[ht!]
\plotone{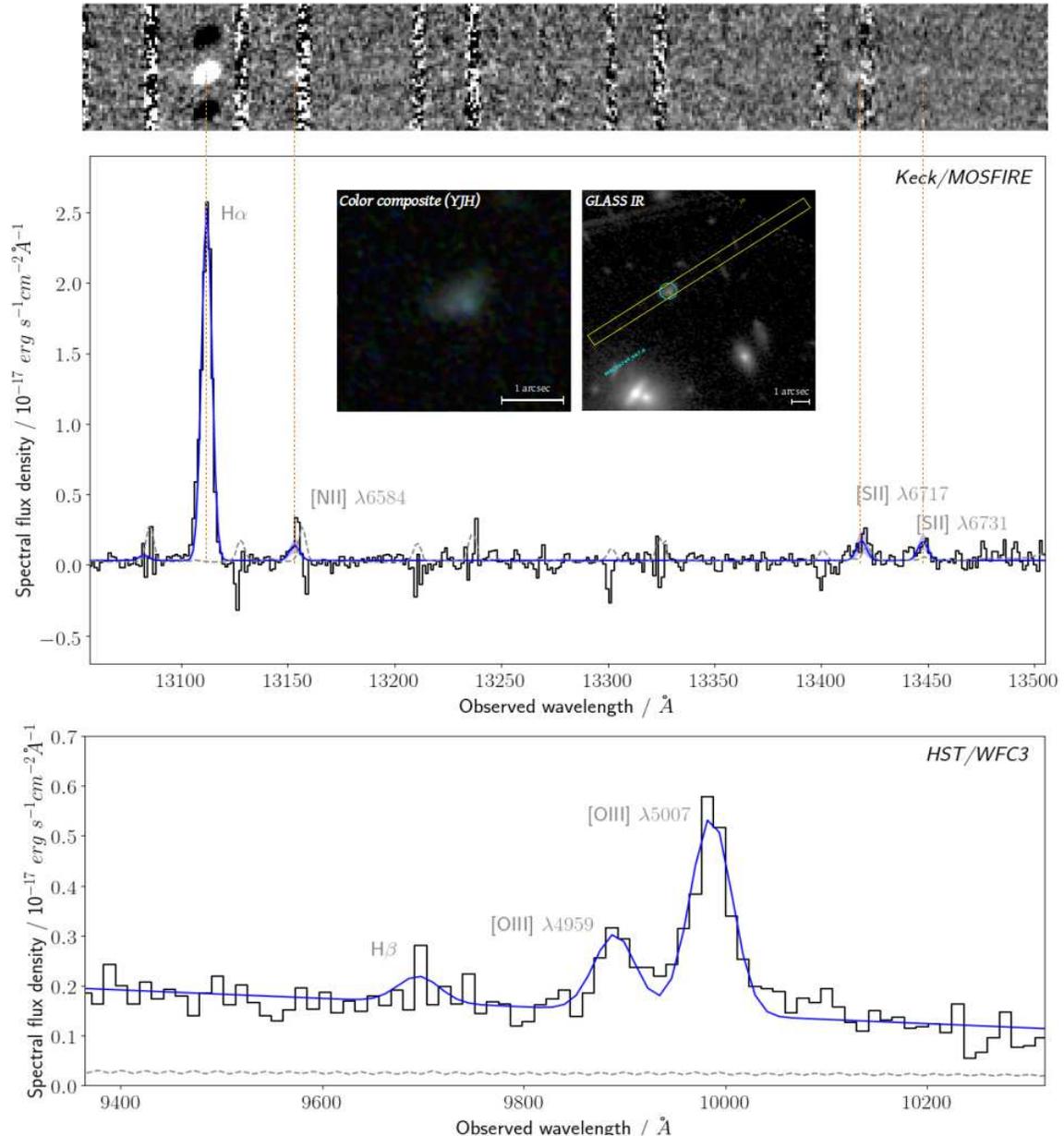}
\caption{\textit{Top:} zoom of area of interest in 2D $J$-band spectrum from MOSFIRE. \textit{Panel 1:} Zoom in on red end of the MOSFIRE $J$-band spectrum of MACS0744\_667.0. Black step-plot shows integrated 1D MOSFIRE signal. Grey dashed line shows 1$\sigma$ uncertainty associated with each wavelength bin. Blue line shows our best-fit spectrum with 2$\sigma$ uncertainty depicted by the lighter blue shaded region. Inset is a color composite from CLASH \emph{HST} $Y$, $J$, and $H$ band imaging as well as the GLASS alignment reference image showing the MOSFIRE mask slit placement. \textit{Panel 2:} Zoom of lines of interest from integrated 1D \emph{HST}/WFC3 $G102$ grism spectrum of MACS0744\_667.0 from the GLASS survey. Color coding is the same as that for panel 1.} \label{fig:spectrum}
\end{figure*}

\section{Analysis} \label{sec:analysis}

\subsection{1D Spectrum Calibration and Construction} \label{sub:reduce_1D}

The sensitivity curve for our observations has been derived from a 1D $J$-band MOSFIRE spectrum of an A5-type reference star (observed with MOSFIRE on 13th June 2016) obtained by integrating over the full width at half maximum (FWHM) spatial extent of the 2D spectrum. We remove the intrinsic stellar spectrum shape by dividing the measured stellar spectrum by an A5-type reference spectrum from the CALSPEC Calibration Database\footnote{\url{http://www.stsci.edu/hst/observatory/crds/calspec.html}}. Thus, the shape of this sensitivity curve gives the response versus wavelength of the set up, due to both the instrument itself and the atmosphere. A faint continuum detection for our object allows us to use its known $J$-band magnitude (from \emph{HST} imaging) to derive a normalizing factor to calibrate the flux in physical units. Applying this sensitivity curve and normalization factor to the 1D signal extracted by integrating the 2D MOSFIRE spectrum along the 1.12 arcsec spatial extent of the \SII signal, we recover our 1D $J$-band spectrum of MACS0744\_667.0, shown in Fig~\ref{fig:spectrum}.

\subsection{Stellar Mass} \label{sub:stellar_mass}

We derive the stellar mass of MACS0744\_667.0 from its spectral energy distribution, following the methods described in our previous analyses of GLASS targets \citep[e.g.][]{Jones15b, Hirtenstein18}. We use 16-band HST photometry spanning observed wavelengths 0.2--1.6 micron from the Cluster Lensing and Supernova Survey with Hubble \citep[CLASH;][]{Postman12}. The contribution of strong emission lines \OIIIs, H$\beta$, H$\alpha$+\NIIs, and \SII given in Table \ref{tab:line_fluxes} are subtracted from the broad-band continuum fluxes. Emission line corrected photometry is then fit with the stellar population synthesis code FAST \citep{Kriek09}. We adopt \citet{BruzualCharlot03} spectral templates with a Chabrier IMF, solar metallicity, \citet{Calzetti00} dust attenuation curve, and an exponentially declining star formation history. Our analysis is relatively insensitive to adopted stellar metallicity. Assuming a sub-solar $Z=0.2 Z_{\odot}$, comparable to the derived gas-phase value, changes the best-fit stellar mass by only 0.01 dex which is negligible compared to the uncertainty.
Since formal statistical photometric uncertainties typically underestimate the total error \citep[e.g.][]{Ilbert06}, we scale the uncertainties in flux by a multiplicative factor such that the best-fitting template has a reduced $\chi^2_{\nu} = 1$. This increases the typical photometric uncertainty from $\sim$0.03 to 0.08 magnitudes. The resulting best-fit stellar mass is $\text{log}(M_*/M_\odot)=8.62^{+0.10}_{-0.06}$ after correcting for lensing magnification (MACS0744 cluster lens modelling obtained from \citealt{Hoag19}). Note that the subtraction of strong emission lines from the broad-band fluxes reduced the best-fit stellar mass by 0.07 dex.

\subsection{Line fitting} \label{sub:line_fits}

Our line fitting procedure was run on a wavelength subset of the full $J$-band spectrum bounded by the bright sky lines at 13055 \AA{} and 13505 \AA{}. Given the continuum in this region is only very tenuously detected, we assume it to be flat at a level taken as the median flux value of the fitting region. After removing this continuum, we obtain line fluxes by fitting a five-peaked Gaussian simultaneously to H$\alpha$, \NII $\lambda$6548, \NII $\lambda$6584 and \SII $\lambda\lambda$6717, 6731 with a $\chi^2$ minimization procedure. To minimize free parameters during fitting, we link all peak centroids and fit only for redshift and assume all peaks have equal line-width. Additionally, the peak height of \NII $\lambda$6548 is assumed to be one third that of \NII $\lambda$6584. Thus we fit for a total of six parameters.

Given the faint \NII and \SII emission lines and their proximity to sky lines, we carried out a robust determination and characterization of the errors affecting these flux measurements. In particular, the red shoulder of the \NII $\lambda$6548, \NII $\lambda$6584 and \SII $\lambda$6717 lines were subject to high RMS error according to the MOSFIRE DRP due to possible sky line contamination.

We estimate our uncertainties with a so-called bootstrapping method by perturbing the one-dimensional calibrated spectrum at each wavelength by drawing from a normal distribution with a standard deviation equal to the RMS error assigned to that wavelength by the MOSFIRE DRP. We perform our line-fitting procedure on 1000 realizations of these perturbed `synthetic' spectra. The mean and standard deviation on the resulting distribution of line-fluxes can be adopted as the measured line flux and its 1$\sigma$ uncertainty.

We find that when applied to the full fitting range, this bootstrapping method tends to over-estimate the flux by around 2$\sigma$ when compared to the standard chi-squared fit. However, if we fit again for our set of emission lines, discarding the potentially contaminated values on the red shoulders of the  \NII $\lambda$6548, \NII $\lambda$6584 and \SII $\lambda$6717 lines ($\lambda$ discarded if $\lambda_\textit{centroid} < \lambda < \lambda_\textit{centroid}+3\times\sigma_{H\alpha}$; less than 20 values discarded in total), both the bootstrapping and standard chi-squared agree within 1 sigma of the lower value from the original chi-squared fit using all of the values (as opposed to the over-estimated bootstrapping mean value). Thus, we adopt the values and the uncertainty from the bootstrapping method as applied to this amended wavelength set without the discarded points. These line fluxes are given in Table~\ref{tab:line_fluxes}.

Upon measuring line fluxes, we verify the flux calibration by comparing with measurements from GLASS HST data, noting that the \Ha and \NII emission lines are blended in the HST spectra. The fluxes are consistent to within $\sim1.5 \sigma$ and the line ratio \SIIs/(H$\alpha$+\NIIs) agrees to within $< 0.1$ dex.

After subtracting instrument dispersion, estimated at $28.7$ km s$^{-1}$ for MOSFIRE $J$-band, we obtain a rest frame \Ha velocity dispersion of $26.7\pm0.5$ km s$^{-1}$.
This includes a natural linewidth of $\sigma_{0} = 3.2$ km s$^{-1}$ for \Has, and thermal broadening estimated conservatively as $\sigma_{th} = 11\pm2$ km s$^{-1}$ (corresponding to nebular temperature $T = 1-2\times10^4$ K; \citealt{GarciaDiaz2008}). Subtracting these effects in quadrature, the intrinsic velocity dispersion of the galaxy is $24.1\pm0.9$ km s$^{-1}$. Given the MOSFIRE slit setting (0\farcs7) and the effective radius of the galaxy (0\farcs3, derived from HST imaging), we do not expect significant systematic error from the good seeing conditions ($\sim$0\farcs4-0\farcs7) during the run as the source image was filling the slit.
In addition, the \Ha emission line width is resolved at high significance compared to sky lines, and the low derived dispersion supports a small dynamical mass. From the half-light radius measured with SourceExtractor \citep{Bertin96}, and following \citet{Erb06b}, we find a dynamical mass of $\text{log}(M_{\text{dyn}}/M_\odot) \approx 9.0$. These dynamical results support the SED-based stellar mass derived in \S \ref{sub:stellar_mass}, with $\text{log}(M_{*}/M_{\text{dyn}}) \approx -0.4$.

\begin{deluxetable}{lC}[b!]
\tablecaption{Fluxes of prominent spectral lines and derived properties of MACS0744\_667.0 \label{tab:line_fluxes}}
\tablecolumns{2}
\tablewidth{0pt}
\tablehead{
\colhead{Spectral line \tablenotemark{a}} &
\colhead{Flux \tablenotemark{b}}
}
\startdata
H$\beta$ & 2.72\pm0.27\\
\OIII $\lambda$4959 & 7.93\pm0.10\\
\OIII $\lambda$5007 & 20.51\pm0.80\\
\hline
\Ha & 14.94\pm0.11\\
\NII $\lambda$6584 & 0.63\pm0.137\\
\SII $\lambda$6717 & 0.84\pm0.228\\
\SII $\lambda$6731 & 0.79\pm0.161\\
\hline
\hline
\multicolumn{2}{c}{Derived Properties}\\
\hline
$z_{\text{MOSFIRE}}$ & 0.997\pm(3\times10^{-6}) \\
$z_{\text{GLASS}}$ & 0.994\pm0002 \\
log($M_* / M_\odot$) & 8.62^{+0.10}_{-0.06} \\
$\sigma_{\text{H}\alpha}$ / (km s$^{-1}$) & 24.1\pm0.5 \\
$12+\text{log}(\text{O}/\text{H})$ & 7.99\pm0.13~ \tablenotemark{c} \\
$n_e$ / (cm$^{-3}$) & \lesssim 1542~ \tablenotemark{d} \\
\enddata
\tablenotetext{a}{H$\beta$ and \OIII lines obtained with \emph{HST}/WFC3 G102 grism spectroscopy from the GLASS survey. Remaining lines from $J$-band MOSFIRE spectroscopy.}
\tablenotetext{b}{Fluxes in units of 10$^{-17}$ erg s$^{-1}$ cm$^{-2}$.}
\tablenotetext{c}{Determined with \citetalias{Dopita16} diagnostic. Quoted uncertainty does not include systematic effects. Refer to \S \ref{sub:metallicity} and \S \ref{sec:discussion} for more information.}
\tablenotetext{d}{Refer to \S \ref{sub:density} for details.}
\end{deluxetable}

\subsection{GLASS Line Fluxes} \label{sub:glass_fitting}

In addition to our MOSFIRE $J$-band observations, we obtained line fluxes for H$\beta$ and [OIII] from the GLASS slitless spectroscopic observations. The 1D grism spectra from GLASS are included in the high-level science products publicly released by the GLASS team, available from STScI/MAST\footnote{https://archive.stsci.edu/prepds/glass/}.

The line fitting procedure follows a process similar to that outlined in Section \ref{sub:line_fits}. We fit over the observed wavelength range 9361 \AA{} $\leq \lambda \leq 10313$ \AA{}, modelling the continuum as a best-fit linear function over this range. We then fit a three-peaked Gaussian profile to the G102 GLASS spectrum (which resolves the [OIII] emission), minimizing free-parameters by fitting for redshift, line-width (assumed equal for all lines), and the areas of each peak.

The low wavelength resolution of the \emph{HST}/WFC3 G102 grism creates difficulties when fitting a continuum, thus uncertainties in the grism line fluxes are likely dominated by uncertainties in the continuum. The uncertainties quoted in table \ref{tab:line_fluxes} are obtained by propagation of the $1\sigma$ values obtained for each fit parameter from the co-variance matrix output by the line-fitting function.

In the context of the BPT diagram we find a very high log(\OIII$\lambda5007$/H$\beta$) ratio, perhaps caused by continuum fitting uncertainties. Although the measured position of MACS0744\_667 on the BPT diagram is broadly consistent with high ionization $z\sim2-3$ galaxies observed by \citet{Strom18}.

The 1D grism spectra from the GLASS data products were flux calibrated independently of this analysis and direct comparison with MOSFIRE line fluxes derived in \S \ref{sub:line_fits}, for example the high apparent Balmer decrement measured ($H\alpha/H\beta=5.6$), may not be reliable.

The redshift fit obtained from the GLASS data ($z_{\text{GLASS}}=0.994$) is slightly offset from that of the MOSFIRE data ($z_{\text{MOSFIRE}}=0.997$). Given the superior wavelength resolution of MOSFIRE, we take $z_{\text{MOSFIRE}}$ to be the source redshift.

\subsection{Metallicity} \label{sub:metallicity}

The suite of measured line fluxes available to us is quite unique for a galaxy at $z\sim1$ with $\text{log}(M_*/M_\odot) \leq9.0$, affording us a range of available metallicity diagnostics. We derive metallicities from diagnostics employing the following line ratios: $\text{N2} = \text{log}(\text{\NII} \lambda6584/\text{\Has})$, $\text{O3N2} = \text{log}((\text{\OIII} \lambda5007/\text{H}\beta)/(\text{\NII} \lambda6584/\text{\Has}))$, and $\text{N2S2H}\alpha = \text{log}(\text{\NII} \lambda6584/\text{\SII} \lambda\lambda6717, 6731) + 0.265\times \text{N2}$. N2 and O3N2 are translated into metallicities using calibrations from \citet{PettiniPagel04} based on a sample of \HII regions with direct ($T_e$) metallicity measurements, while metallicity is inferred from N2S2 using the \citet{Dopita16} diagnostic, based on theoretical models. These calibrations are as follows:

\begin{equation}
    12+\text{log}(\text{O}/\text{H})=8.90+0.57\times\text{N2}
\end{equation}
\begin{equation}
    12+\text{log}(\text{O}/\text{H})=8.73-0.32\times\text{O3N2}
\end{equation}
\begin{equation}
    12+\text{log}(\text{O}/\text{H})=8.77+\text{N2S2H}\alpha.
\end{equation}

Applying these diagnostics to our measured line ratios yields values of $Z_{\text{N2}} =8.11\pm0.05, Z_{\text{O3N2}} =8.01\pm0.03$ and $Z_{\text{D16}}=7.99\pm0.13$, where $Z=12+\text{log}(\text{O}/\text{H})$, as given in Table~\ref{tab:Zs}.
Uncertainties quoted here are strictly measurement uncertainties; systematic uncertainties are discussed in \S\ref{sec:discussion}.

\begin{deluxetable}{lcC}[b!]
\tablecaption{Metallicity as derived by different available diagnostics. Quoted uncertainties do not include systematic effects. \label{tab:Zs}}
\tablecolumns{3}
\tablenum{2}
\tablewidth{0pt}
\tablehead{
\colhead{Line ratios\tablenotemark{$\dagger$}} &
\colhead{Calibration reference} &
\colhead{$12+\text{log}(\text{O}/\text{H})$}
}
\startdata
N2 & \citet{PettiniPagel04} & 8.11\pm0.05 \\
O3N2 & \citet{PettiniPagel04} & 8.01\pm0.03 \\
N2S2H$\alpha$ & \citet{Dopita16} & 7.99\pm0.13 \\
\enddata
\tablenotetext{\dagger}{Definitions of listed ratio names given in \S \ref{sub:metallicity}}
\end{deluxetable}

\subsection{Electron Density} \label{sub:density}

The ratio between the \SII $\lambda$6717 and \SII$\lambda$6731 in the \SII doublet is the most widely used measure of electron density in \HII regions. Our detection of this doublet allows us to put constraints on the electron density in this target. We calculate a ratio of \SII $\lambda$6717/\SII$\lambda$6731 $=1.06\pm0.36$. According to the calibration provided by \citet{ProxaufDENSITY} this places a 1$\sigma$ upper limit on the density of $n_e \leq 1542$ cm$^{-3}$.

\begin{figure*}
    \centering

    \includegraphics[width=0.49\linewidth]{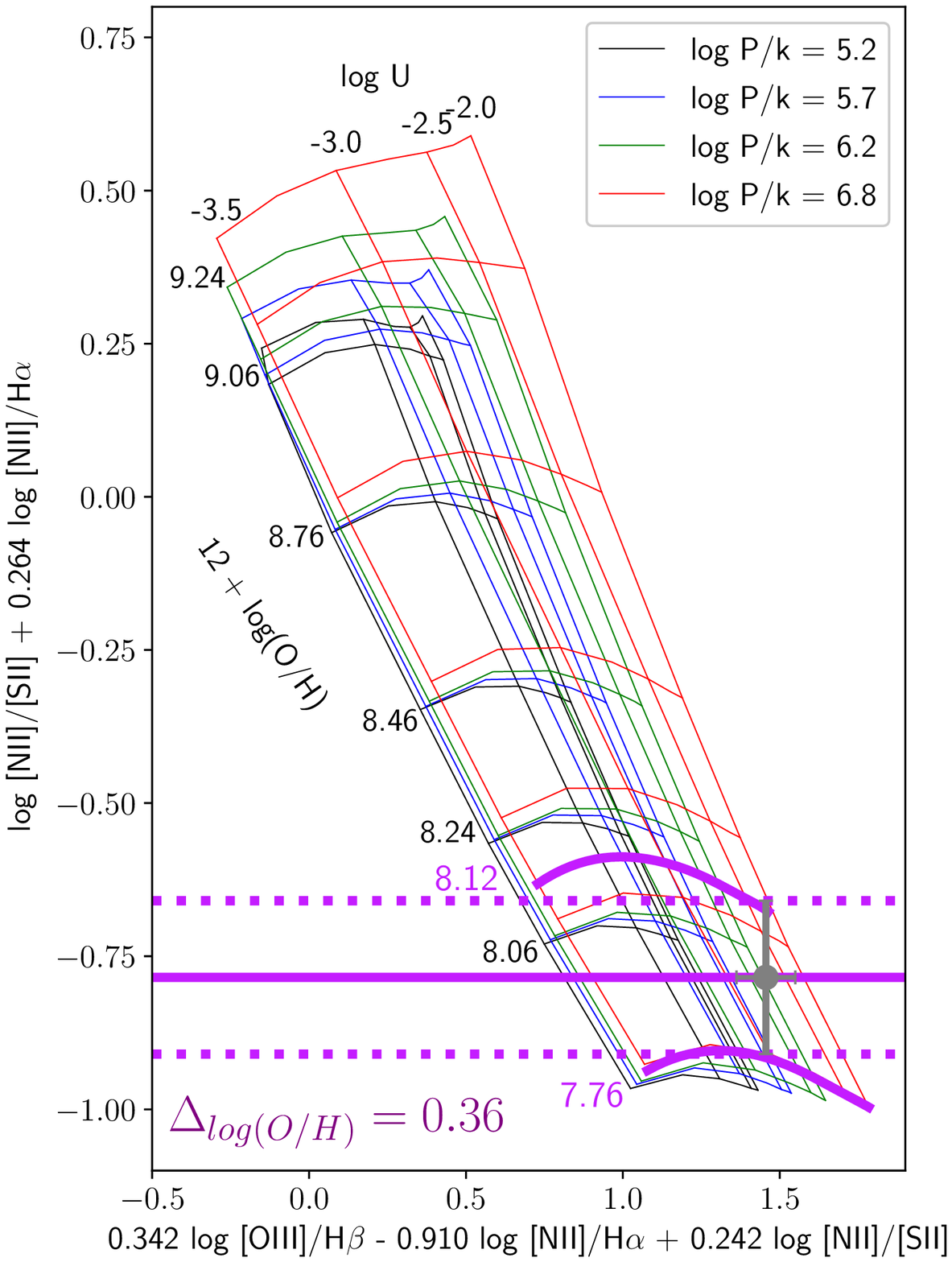}
    \hfill
    \includegraphics[width=0.49\linewidth]{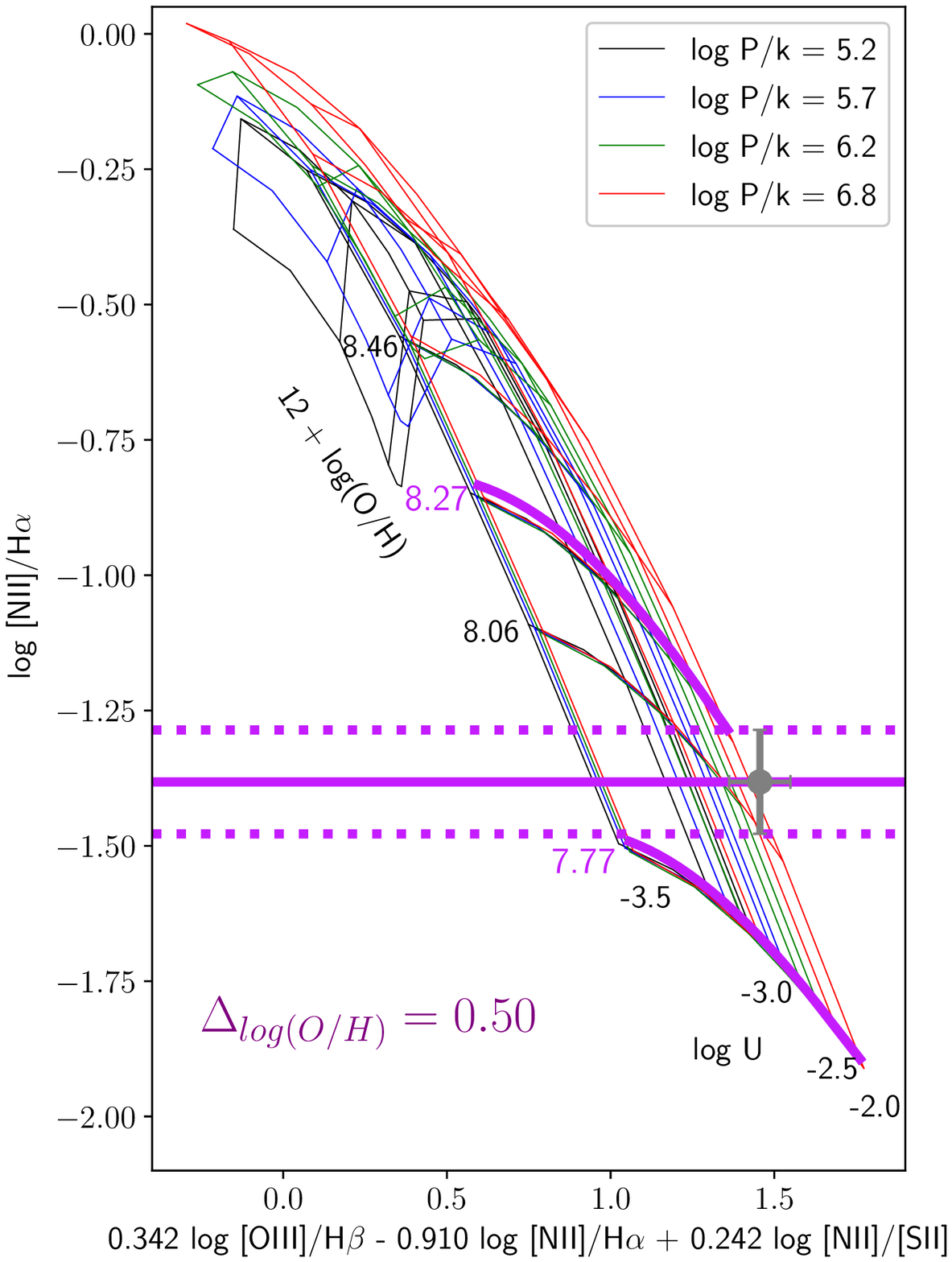}

    \caption{Two projections of theoretical grids (obtained from Dopita, private communication) showing the improved systematic uncertainty afforded by the \citetalias{Dopita16} diagnostic over the classic N2 diagnostic. The grid lines depict metallicity (roughly horizontal) and ionization parameter (steep diagonal) with different grid colors corresponding to different ISM pressures. The vertical axes in each panel are the set of line ratios required by each respective diagnostic (\citetalias{Dopita16} on the left). The horizontal axis is a combination of line ratios that corresponds neatly with ionization parameter with the \citetalias{Dopita16} y-axis at fixed metallicity and pressure; this is convenient for visual clarity. In both cases the grey point represents measurements reported here for MACS0744\_667.0. The horizontal purple lines emphasize the 1$\sigma$ measurement uncertainty obtained for each quantity given by the vertical axes. These uncertainty intervals clearly indicate the improvement of the \citetalias{Dopita16} diagnostic over the simpler N2 diagnostic. Although the D16 line ratio value itself is not as well constrained (due to lower S/N of the \NIIs/\SII ratio), the reduced systematic uncertainty allows for tighter constraints to be placed on the actual metallicity of this object (1$\sigma$ upper and lower bounds depicted by purple contours). A first-order approximation with these grids suggests that knowledge of the N2 ratio alone is unable to constrain metallicity tighter than $8.27 \leq 12+\text{log(O/H)} \leq 7.77$ while the \citetalias{Dopita16} diagnostic constrains the metallicity to within $8.12 \leq 12+\text{log(O/H)} \leq 7.76$ at the 1$\sigma$ level.} \label{fig:grids}
\end{figure*}

\section{Discussion} \label{sec:discussion}

Strong-line methods are currently the only feasible route to metallicity studies with large samples at high redshifts, particularly for low-mass galaxies.
In addition to measurement uncertainty, present at some level in any observation, strong-line measurements in particular suffer from systematic uncertainties caused by degeneracy between metallicity and other physical parameters (ionization parameter, N/O abundance ratio, etc) on the line ratios being employed by the diagnostic.
These uncertainties can arise from both stochastic variations of these physical parameters, as well as any systematic variations that may be present in the high-redshift universe.
Understanding and minimizing the uncertainties associated with these methods is therefore a critical and open issue.

As a result of the observational challenges associated with assembling large suites of emission lines for low-mass galaxies, the \NIIs/\Ha diagnostic (N2; \citealt{PettiniPagel04}) is widely used in high redshift studies as the required lines are relatively strong and are close in wavelength \citep[e.g.][]{Erb06, Wuyts12, Yabe14}. By contrast, many widely used strong-line diagnostics, such as O3N2 \citep{PettiniPagel04} or $R_{23}$ \citep{Zaritsky94}, can deliver lower systematic uncertainty by better accounting for degeneracy with other physical parameters, but are considerably more challenging to obtain at high redshifts due to the wavelength coverage required. However, the reliance of N2 on a single line ratio makes it vulnerable to systematic uncertainties; in particular those caused by degeneracy with ionization parameter and N/O abundance ratio.

\subsection{Ionization Parameter Dependence}

\citet{Strom18} show that ionization parameter is the physical parameter to which nebular spectra respond most sensitively. Accordingly, the dependence of the N2 ratio on ionization parameter introduces systematic uncertainty into metallicities derived from N2.
The detection of the \SII doublet in this object allows us to adopt the \citetalias{Dopita16} approach of incorporating \NIIs/\SII in addition to N2, which drastically reduces degeneracy between metallicity, ionization parameter and ISM pressure.

Grids obtained from M. Dopita (private communication) plotted in Figure \ref{fig:grids} clearly highlight the short-comings of the classic N2 diagnostic. The grids show how the expected line ratios vary with ionization parameter and ISM pressure at fixed metallicity. We plot MACS0744\_667.0 as reported in this letter onto these grids to illustrate the tighter constraint afforded by using the \citetalias{Dopita16} diagnostic. Although the line ratio itself has a larger uncertainty, the vastly reduced ionization dependence leads to an overall better metallicity measurement (constrained to a range of $0.36$ dex for \citetalias{Dopita16} c.f. $0.50$ for N2; 1$\sigma$ upper and lower bounds indicated by the purple curves in Fig~\ref{fig:grids}). Furthermore, unlike systematic uncertainty, measurement uncertainty reduces with higher signal-to-noise, meaning prospects for placing tight constraints on the metallicities of low-mass galaxies at high-redshift are greatly improved for \citetalias{Dopita16} compared to when the N2 line ratio is used in isolation. This approach is made feasible in this low-mass and high-redshift regime by virtue of the spectral proximity of the \SII doublet to the \Ha and \NII$\lambda$6584 lines, meaning they can be obtained without having to essentially double time-on-target requirements by observing with additional filters.

\begin{figure*}
    \centering
    \includegraphics[width=1\linewidth]{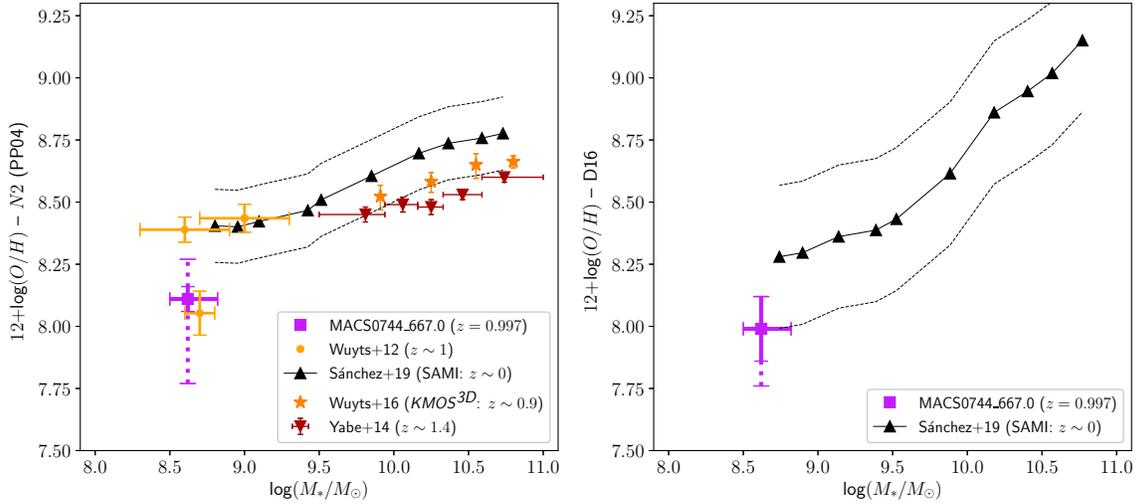}
    \caption{
    MACS0744\_667.0, shown as the purple square, resides in a region of $M_* - \text{log}(O/H)$ space that is sparsely populated by existing observations. Solid error bars show $1\sigma$ measurement uncertainty while the dotted error bars depict our estimate of additional systematic uncertainty due to uncertainty in the ionization parameter (see Fig~\ref{fig:grids}).
    \textit{Left panel:} Mass-metallicity relation with metallicities derived from the N2 diagnostic \citep{PettiniPagel04}.
    Orange stars are metallicities derived from N2 ratios of stacks of spectra at $0.6<z<1.1$ binned by mass from \textit{KMOS}$^\textit{3D}$ \citep{Wuyts16}. Orange circles are N2 metallicities from individual lensed objects at $z\sim1$ from \citet{Wuyts12}. Red inverted triangles are N2 metallicities of $z\sim1.4$ galaxies binned by mass from \citet{Yabe14}. Displayed error bars reflect only measurement uncertainty for these objects.
    Black triangles show the N2 gas-phase metallicities binned by mass measured at 1 $R_e$ in SAMI galaxies at $z\sim0$ with dashed black lines indicating the average residual 1$\sigma$ scatter after fitting a Mass-Metallicity Relation \citep{Sanchez19}.
    Some evolution is seen from $z\sim1.4$ to $z\sim0$ above log$(M_*/M_\odot)\geq9.5$, however neither large sample at high redshift is able to probe below log$(M_*/M_\odot)<9.5$ where evolution is expected to be most significant. Despite the low measurement uncertainty associated with MACS0744\_667.0 the large systematic uncertainty limits the degree to which insight can be gained from this measurement.
    \textit{Right panel:}  Mass-metallicity relation with metallicities derived from the \citetalias{Dopita16} diagnostic.
    Black triangles show the \citetalias{Dopita16} gas-phase metallicities measured at 1 $R_e$ in SAMI galaxies at $z\sim0$ \citep{Sanchez19}. Here, MACS0744\_667.0 has a larger measurement uncertainty, however the additional systematic uncertainty is less problematic, suggesting larger samples using this diagnostic will be more effective in constraining redshift evolution of the mass-metallicity relation.
    } \label{fig:MZR}
\end{figure*}

Figure~\ref{fig:MZR} shows metallicity, including estimated systematic uncertainty due to ionization parameter (dotted error bars), plotted against the stellar mass (see \S~\ref{sub:stellar_mass}) for this object and a few available in the literature in a similar redshift range ($z\sim1$). The black triangles in Fig~\ref{fig:MZR} show the $z\sim0$ MZR derived from the SAMI survey \citep{Sanchez19}. Given that chemical evolution of low-mass galaxies is expected to be more significant at later times \citep[e.g.][]{Henry13b}, samples of objects with stellar masses below $\text{log}(M_*/M_\odot)<9.0$ beyond redshift $z\gtrsim1$ promise to provide valuable insight into evolutionary processes driving the galaxy population. In Fig~\ref{fig:MZR} the limitations caused by systematic uncertainties can be seen. Despite the low measurement uncertainty for our N2 metallicity (left panel), the large systematic uncertainty means the resultant measurement provides little insight to distinguish between cases where significant or very little evolution occurs from $z\sim1$ to $z\sim0$.

\subsection{Nitrogen-to-Oxygen Ratio Dependence}

A remaining concern, however, is that the N2 ratio is primarily sensitive to the nitrogen abundance. Thus, oxygen abundances can only be inferred using some (implicitly or explicitly) assumed N/O ratio. Systematic uncertainty on measurements of oxygen abundances conducted in this way can therefore be introduced in two main ways.
First, even assuming an appropriate N/H-to-O/H conversion can be applied, stochastic variations of log(N/O) at fixed log(O/H) add to the overall uncertainty of the final metallicity measurement.
In a sample of objects with log(N/O) and 12+log(O/H) direct measurements \citep{Berg12, Pilyugin12} we found that among the objects with $7.8<12+\text{log}(O/H)<8.2$, the standard deviation in log(N/O) was $\sigma_{\text{log(N/O)}}=0.13$ dex, comparable to line ratio measurement uncertainties. Although strong-line measurements at $z\sim2.3$ by \citet{Strom18} suggest this scatter could be as large as 0.8 dex in high-redshift galaxies. This stochastic variation limits the accuracy of individual metallicity measurements; larger samples are required to account for this effect.
Second, some authors suggest that the N/O abundance ratio has undergone evolution with redshift \citep[e.g.][]{Masters14}. Although, \citet{Steidel16} suggest that $z\sim2$ galaxies on average lie within the same trend as local galaxies. This will critically affect metallicities derived using nitrogen lines. However, it is a difficult issue to address, requiring large samples of high quality spectra at high redshift. If, indeed, high-redshift galaxies do exhibit different N/O ratios to local galaxies the assumed N/O parameterization may be inappropriate. Modest samples of ``direct'' $T_e$ measurements at $z>1$ may help to further understand this.

\subsection{Future Prospects}

While better constraints on the Mass-Metallicity relation below log$(M_*/M_\odot) \leq 9.0$ at $z\gtrsim1$ promise unique insights into evolution of the galaxy population, progress has been limited by the associated observational challenges.
Strong-line metallicity measurements are the only feasible approach to making progress in this area, however the systematic uncertainties associated with the simple N2 diagnostic clearly limit its effectiveness in distinguishing between different evolutionary processes.

An additional source of uncertainty not discussed here is the contribution to the measured \SIIs/H$\alpha$ ratio from diffuse ionized gas (DIG). The consequence of this is that global metallicities derived from N2S2H$\alpha$ will be sensitive to variations in the fraction of DIG (f$_{DIG}$) in the galaxy population. As highlighted in \citet{Shapley19}, if high-redshift galaxies follow the same relation between f$_{DIG}$ and $\Sigma_{SFR}$ as local galaxies, N2S2H$\alpha$ would vary systematically with redshift and thus not be appropriate for comparing high- and low-redshift samples. Thus, further observations are required to determine the degree to which variations in f$_{DIG}$ would affect the systematic uncertainties induced in a sample at fixed redshift beyond $z\gtrsim1$.

Approaches that better constrain degeneracy by including more emission line ratios are certainly preferable, although challenging in this regime where long integration times are required to accurately measure even the strongest metal emission lines.
Approaches that use photoionization models to simultaneously fit for all of these physical parameters including metallicity \citep[e.g.][]{PerezMontero14, Blanc15} appear to lend themselves naturally to this context, however we did not include these in this discussion as the low measured S/N for MACS0744\_667.0 implies that both the specific details of the input model adopted, and parameter priors could have a substantial impact on the inference.
While future facilities will certainly aid progress in this area, applying the \citetalias{Dopita16} diagnostic to deep observations with existing instruments can improve the ionization dependence of existing N2-based constraints without requiring the factor of 2-3 increase in time-on-target associated with many other strong-line methods.

\section{Conclusion} \label{sec:conclusion}

Extending constraints on the high-redshift ($z\gtrsim1$) Mass-Metallicity Relation to masses below log$(M_*/M_\odot)<9.0$ promises powerful insight into the evolutionary processes that govern the galaxy population.
Currently strong-line methods are the only viable approach for expanding metallicity measurements to high-redshift dwarf galaxies.

However, particularly when the number of available emission lines is small, metallicity measurements made with strong-line methods may suffer from degeneracy with other physical parameters such as ionization parameter,  chemical abundance ratios and ISM pressure. Derived metallicities can be affected by either stochastic or systematic variations in these properties among the high-redshift galaxy population.

In this contribution we have presented MOSFIRE $J$-band spectroscopy of MACS0744\_667.0, a low mass ($\text{log}(M_*/M_\odot)=8.62^{+0.10}_{-0.06}$) star-forming galaxy at redshift $z=0.997$ magnified $1.4\times$ by CLASH cluster MACS0744 in which we observe detection of \NII $\lambda$6584 and \SII $\lambda\lambda6717, 6731$ alongside strong \Ha detection. Additionally, we derive H$\beta$, \OIII $\lambda$4959 and \OIII $\lambda$5007 line fluxes from HST/WFC3 G102 grism spectroscopy from the GLASS data release. Access to this set of emission lines is quite unique for a galaxy of this mass at this redshift.

We derive metallicity from N2 ($12+\text{log}(\text{O}/\text{H})=8.11\pm0.05$  with statistical uncertainty; $12+\text{log}(\text{O}/\text{H})=8.11^{+0.16}_{-0.34}$ including additional systematic uncertainty) as well as N2S2H$\alpha$ ($12+\text{log}(\text{O}/\text{H})=7.99\pm0.13$; $12+\text{log}(\text{O}/\text{H})=7.99^{+0.13}_{-0.23}$). While the inclusion of the \NIIs/\SII ratio in N2S2H$\alpha$ increases the measurement uncertainty, we find that even in this case where S/N$_{N2S2}$ is small, the improved mitigation of the dependency on ionization parameter reduces the overall uncertainty on the metallicity measurement. We estimate that uncertainty due to N/O abundance ratio is likely of comparable order to the measurement uncertainty of N2S2H$\alpha$. Large samples of high quality spectra of high-redshift dwarf galaxies are needed to assess if there is a systematic variation of this abundance ratio at high-redshift and at what level the stochastic variation impacts dwarf galaxy metallicities.

Further progress in extending the high-redshift Mass-Metallicity Relation to dwarf galaxies requires careful consideration of the global trends among dwarf galaxies in all physical parameters, not just metallicity. Deeper surveys targeting low-mass objects at $z\gtrsim1$ employing existing multiplexed NIR instruments (e.g. \emph{Keck}/MOSFIRE or \emph{VLT}/KMOS) will improve understanding of the stochastic variations in these properties among the high-redshift dwarf population, providing unique insight into the evolutionary processes that govern the galaxy population.


\acknowledgments{}

We are grateful to the late M. A. Dopita for providing us with the theoretical grids presented in \citet{Dopita16} and would like to acknowledge his prolific and extensive contributions to the field spanning many decades. This research was supported by the Australian Research Council Centre of Excellence for All Sky Astrophysics in 3 Dimensions (ASTRO 3D), through project number CE170100013. AJC acknowledges support from an Australian Government Research Training Program (RTP) Scholarship.

%

\vspace{5mm}
\facility{Keck(MOSFIRE)}

\bibliographystyle{aasjournal}
\bibliography{mosfire}




\end{document}